\documentclass[12pt,preprint]{aastex}
\def\jcap{JCAP}
\def\beq{\begin{equation}}
\def\eeq{\end{equation}}
\def\ben{\begin{eqnarray}}
\def\een{\end{eqnarray}}
\def\num{\sum m_{\nu}}
\def\nus{\texttt{MassiveNuS}}
\def\nulcdm{\nu\Lambda{\rm CDM}}
\def\ev{\,{\rm eV}}
\def\munit{\,h^{-1}\! M_{\odot}}

\def\db{D_{\rm B}}

\shortauthors{Ryu \& Lee}
\begin{document}
\title{The Splashback Mass Function in the Presence of Massive Neutrinos}
\author{Suho Ryu and Jounghun Lee}
\affil{Astronomy program, Department of Physics and Astronomy,
Seoul National University, Seoul  08826, Republic of Korea \\
\email{shryu@astro.snu.ac.kr, jounghun@astro.snu.ac.kr}}
%%%%%%%%%%%%%%%%%%%%%%%%%%%%%%%%%%%%%%%%%%%%%%%%%%%%%%%%%%%%%
\begin{abstract}
We present a complementary methodology to constrain the total neutrino mass, $\num$, based on the diffusion coefficient of the splashback mass function of dark 
matter halos.  Analyzing the snapshot data from the Massive Neutrino Simulations, we numerically obtain the number densities of distinct halos identified via the 
SPARTA code as a function of their splashback masses at various redshifts for two different cases of $\num=0.0\ev$ and $0.1\ev$. 
Then, we fit the numerical results to the recently developed analytic formula characterized by the diffusion coefficient that quantifies the degree of ambiguity in the 
identification of the splashback boundaries. 
Our analysis confirms that the analytic formula works excellently even in the presence of neutrinos and that the decrement of its diffusion coefficient with redshift 
is well described by a linear fit, $B(z-z_{c})$, in the redshift range of $0.2\le z\le 2$. It turns out that the massive neutrino case yields significantly lower value of $B$ 
and substantially higher value of $z_{c}$ than the massless neutrino case, which indicates that the higher masses the neutrinos have, the more severely the splashback 
boundaries become disturbed by the surroundings. Given our result, we conclude that the total neutrino mass can in principle be constrained 
by measuring how rapidly the diffusion coefficient of the splashback mass function diminishes with redshifts at $z\ge 0.2$. We also discuss the anomalous behavior 
of the diffusion coefficient found at lower redshifts for both of the $\num$ cases, and ascribe it to the fundamental limitation of the SPARTA code at $z\le 0.13$.
\end{abstract}
%%%%%%%%%%%%%%%%%%%%%%%%%%%%%%%%%%%%%%%%%%%%%%%%%%%%%%%%%%%%%

\section{Introduction}\label{sec:intro}

The dark matter halos form through gravitational collapse of the initially overdense sites. As the gravitational collapse is a highly nonlinear 
process occurring after the shell crossing, the linear or even higher-order perturbation theory is not eligible for the analytical description of the halo 
formation.  Only for the extreme case that the collapse of an isolated overdense site occurs completely in a spherically symmetry way, the halo formation 
can be analytically described by the top-hat spherical dynamics, which basically predicts that the virialized halos form from the sites whose linearly extrapolated 
density contrast exceeds a certain value, called the virial density threshold \citep{GG72}.  A unique aspect of the top-hat spherical dynamics is that the virial 
density threshold is independent of halo mass and quite insensitive to the initial conditions of the universe \citep[e.g.,][]{PS74,lah-etal91,eke-etal96,pac-etal10}.

In realistic cases where the overdense sites are usually not isolated and the gravitational collapse does not proceed in a spherically symmetric way,  
the simple analytical description of the halo formation in terms of a constant  value of the virial density threshold is no longer attainable 
\citep{BM96,ST99,CL01,SMT01,tin-etal08}. 
Nevertheless, several theoretical works based on $N$-body simulations indicated that the condition for the realistic formation of dark matter halos 
can still be expressed in terms of the virial density threshold by treating it as a mass-dependent stochastic variable rather than a deterministic constant 
value \citep[e.g.,][]{rob-etal09,MR10a,MR10b,CA11a,CA11b}. The mass-dependence of the mean value of the virial density threshold stems from the departure of the 
gravitational collapse from the top-hat spherical dynamics, while its stochastic nature reflects the ambiguity in the identification of non-isolated halos. 

In the high-mass limit corresponding to the massive cluster halos, the virial density threshold becomes less non-spherical but more stochastic 
\citep[e.g.,][]{rob-etal09,MR10a,MR10b}. 
Since the gravitational collapse of highest density peaks occur more or less in a spherically symmetric way \citep{ber94}, the mean value of the 
virial density threshold converges to the spherical value, $\delta_{sc}=1.686$, \citep{GG72}. 
On the other hand, the massive cluster halos often reside at the nodes of cosmic filaments and thus are more vulnerable to the disturbance 
from the surroundings. In consequence, it becomes more ambiguous to identify their clear-cut boundaries, which results in a higher degree of 
stochasticity of the virial density threshold in this limit \citep{MR10b}. 

Very recently, it has been suggested that the ambiguity in the cluster identification can be considerably reduced by using the splashback radius as a physical 
boundary instead of the virial counterpart \citep{RL21}. The splashback radius of a DM halos was originally found in $N$-body simulations as its outer limit 
where the density profile exhibits an abrupt drop \citep{wan-etal09,wet-etal14,DK14,adh-etal14,die21}. 
It corresponds to the radial distance to the apocenter of the elliptical orbit of the latest infalling particle  \citep{adh-etal14,mor-etal15}, 
distinguishing between infalling and orbiting particles, and naturally taking into account the fly-by objects \citep{mor-etal15}. 
If the halo boundary is defined by its splashback radius, which is in fact a solution to the self-similar spherical infall model \citep{FG84,ber85}, 
the unrealistic assumption of {\it isolated} overdense region is no longer required to describe the halo formation, since the splashback boundary generically 
incorporates the pseudo-evolution of halo mass  \citep{die-etal13}.  

Noting that the self-similar spherical infall model gives a lower mean value of the splashback density threshold \citep{sha-etal99}, \citet{RL21} substituted 
it for the virial counterpart in the the generalized excursion set formalism \citep{MR10a,MR10b,CA11a,CA11b} and proposed an analytic formula for the splashback 
mass function characterized by a single parameter, {\it diffusion coefficient}, which measures the degree of the stochasticity of the splashback density 
threshold. Testing the analytically derived splashback mass functions against the $N$-body results for the 
Planck and WMAP7 cosmologies \citep{planck14,wmap7}, 
\citet{RL21} verified its validity in a wide mass range up to the redshift $z\sim 3$. Showing that the redshift evolution of the diffusion coefficient 
sensitively depends on the matter density parameter $\Omega_{m}$, \citet{RL21} suggested that the diffusion coefficient of the splashback 
mass function of dark matter halos be a sensitive probe of the initial conditions of the universe. 

In this paper, we explore if the diffusion coefficient of the splashback mass function can effectively constrain the total neutrino mass, $\num$, speculating 
that the degree of the stochasticity of the splashback density threshold may be sensitive not only to the amount of dark matter but also to its nature. 
The upcoming Sections contain the following contents: a brief review of the analytic formula for the splashback mass function (Section \ref{sec:review}); 
a comparison between the analytic formula and the numerically obtained splashback mass function in the presence of neutrinos and 
a difference in the redshift evolution of the diffusion coefficient between the massless and massive neutrino cases 
(Section \ref{sec:test}); a summary and conclusion (Section \ref{sec:sum}).
Throughout this paper, we will assume a cosmology where the neutrino ($\nu$) is present, and the cosmological constant ($\Lambda$) and cold dark matter 
(CDM) are dominant, i.e., $\nu\Lambda$CDM cosmology.

\section{A brief review of the analytic model}\label{sec:review}

In the original excursion set formalism \citep{PS74,bon-etal91}, the virial mass function, $dN(M,z)/d\ln M$, defined as the number densities of DM halos 
as a function of its logarithmic virial mass, $\ln M$, at redshift $z$, can be analytically written as 
%%%%%%%%%%%%%%%%%%%%%%%%%%%%%%%%%%%%%%%%%%%%%%%%%%%%%%%%%%%%%%%%%%%%%%%%%%%
\beq
\label{eqn:massft}
\frac{dN(M, z)}{d\ln M} =  \frac{\bar{\rho}}{M}\Bigg{\vert}\frac{d\ln\sigma^{-1}}{d\ln M}\Bigg{\vert}f(\sigma)\, , 
\eeq
%%%%%%%%%%%%%%%%%%%%%%%%%%%%%%%%%%%%%%%%%%%%%%%%%%%%%%%%%%%%%%%%%%%%%%%%%%%%
where $\bar{\rho}$ is the mean mass density of the universe, $\sigma(M,z)$ is the rms fluctuation of the linear density field on the mass scale 
of $M$ at redshift $z$, and $f(\sigma)$ is the multiplicity function that counts the number of the initial density peaks whose linearly extrapolated 
density contrast exceeds some {\it virial density threshold} when the rms density fluctuation has the value of $\sigma$. The excursion set theory relates 
$dN/d\ln M$ to the background cosmology through $\sigma(M,z)$ that is often computed as $\sigma^{2}(M,z)\equiv (2\pi^{2})^{-1}\int\,dk\,k^{2}P(k,z)W^{2}(k,M)$
where $P(k,z)$ is the linear density power spectrum and $W(k,M)$ is the top-hat window function on the virial mass scale $M$ \citep{PS74,bon-etal91}.  

This original excursion set theory was generalized by \citet[][2010b]{MR10a} (hereafter, Maggiore-Riotto) who treated the virial density threshold as a 
stochastic random variable whose mean value is $\delta_{sc}=1.686$ according to the top-hat spherical model \citep{GG72}. 
The resulting Maggiore-Riotto model turned out to describe very well the virial mass function of cluster halos, despite that it has only one fitting parameter, 
called the diffusion coefficient, which characterizes the degree of the stochasticity of the virial density threshold.  Noting the limited validity of the Maggiore-Riotto 
formula to the cluster mass scale, \citet[][2011b]{CA11a} modified it by taking into account possible variation of the mean value of the virial density threshold
with mass scale. The  modified formula, which has one additional parameter called the drifting average coefficient, was shown to improve the agreements with 
the numerical results on the galactic mass scale.  

It was \citet{RL21} who for the first time proposed that the generalized excursion set theory can also be used for the evaluation of the splashback mass function. 
From here on, we let $M$ denote not the virial but the splashback mass of a DM halo.  They put forth an idea that the splashback density threshold must be much less 
stochastic than the virial counterpart and that its mean value would not change with the mass scale. Then, they claimed that the splashback multiplicity function could be 
described by the following single-parameter formula, which is the same as the Maggiore-Riotto model but with $\delta_{sc}=1.686$ replaced by $\delta_{sp}=1.52$, the mean 
splashback density threshold predicted in the self-similar spherical infall model \citep{sha-etal99}:
%%%%%%%%%%%%%%%%%%%%%%%%%%%%%%%%%%%%%%%%%%%%%%%%%%%%%%%%%%%%%%%%%%%%%%%%%%% 
%%%%%%%%%%%%%%%%%%%%%%%%%%%%%%%%%%%%%%%%%%%%%%%%%%%%%%%%%%%%%%%%%%%%%%%%%%%
\ben
f(\sigma ; D_{B}) &=& \frac{\delta_{sp}}{\sigma}\sqrt{\frac{2}{\pi(1+D_{B})}} \Biggl\{ \left[1-\frac{\kappa}{(1+D_{B})}\right]
\exp\left[-\frac{\delta_{sp}^2}{2\sigma^2(1+D_{B})}\right]  \nonumber \\
&&+\frac{\kappa}{2(1+D_B)}\Gamma\left[0,\frac{\delta_{sp}^2}{2 \sigma^2(1+D_{B})}\right]\Biggl\}\, .
\label{eqn:multi}
\een
%%%%%%%%%%%%%%%%%%%%%%%%%%%%%%%%%%%%%%%%%%%%%%%%%%%%%%%%%%%%%%%%%%%%%%%%%%%
Here, $\Gamma$ is the incomplete gamma function, $\db$ is the diffusion coefficient defined as the ratio of the variance of the splashback density threshold 
to the mass variance, and $\kappa$ is related to the cross correlation of the linear density field on two different splashback mass scales 
$M$ and $M^{\prime}$ as\footnote{In the work of \citet{RL21}, they adopted the same approximation of $\kappa\approx 0.475$ as used 
in \citet{CA11b}, assuming a $\Lambda$CDM cosmology. However, in the current work, we rigorously calculate $\kappa$ by Equation (\ref{eqn:kappa}).}: 
%%%%%%%%%%%%%%%%%%%%%%%%%%%%%%%%%%%%%%%%%%%%%%%%%%%%%%%%%%%%%%%%%%%%%%%%%%
\begin{equation}
\label{eqn:kappa}
\kappa \approx \frac{\sigma^{2}(M)}{\sigma^{2}(M^{\prime})\left[\sigma^{2}(M)-\sigma^{2}(M^{\prime})\right]}
\left[\langle\delta(M)\delta(M^{\prime})\rangle - \sigma^{2}(M^{\prime})\right]\, .
\end{equation}
%%%%%%%%%%%%%%%%%%%%%%%%%%%%%%%%%%%%%%%%%%%%%%%%%%%%%%%%%%%%%%%%%%%%%%%%%%% 

Comparing Equations (\ref{eqn:massft})-(\ref{eqn:kappa}) with the numerical results from the Erebos $N$-body experiment \citep{DK14,DK15} and 
determining $\db$ as a function of redshift, \citet{RL21} found the following. 
First, despite that it has only one parameter, the analytic formula excellently works in the splashback mass range of  $5\times 10^{12}\le M/(h^{-1}\,M_{\odot})\le 10^{15}$ 
up to $z\sim 3$ for two different $\Lambda$CDM cosmologies. 
Second, the splashback mass function is characterized by a much lower value of $\db$ than the virial mass function, which justifies the assumption that the splashback 
density threshold is less stochastic than the virial counterpart. 

Third, the diffusion coefficient almost monotonically decreases with $z$, well approximated by a linear fit, and vanishing at some critical redshift, $z_{c}$. 
This result implied that the splashback mass function at $z\ge z_{c}$ can be described by the following purely analytic formula with no free parameter: 
%%%%%%%%%%%%%%%%%%%%%%%%%%%%%%%%%%%%%%%%%%%%%%%%%%%%%%%%%%%%%%%%%%%%%%%%%%
\begin{equation}
\label{eqn:para_free}
\frac{dN(M, z)}{d\ln M} =  \frac{\bar{\rho}}{M}\Bigg{\vert}\frac{d\ln\sigma^{-1}}{d\ln M}\Bigg{\vert}\frac{\delta_{sp}}{\sigma}
\sqrt{\frac{2}{\pi}}\Bigg{\{}\left(1-\kappa\right)
\exp\left(-\frac{\delta_{sp}^2}{2\sigma^2}\right)+\frac{\kappa}{2}\Gamma\left[0,\frac{\delta_{sp}^2}{2 \sigma^2}\right]\Bigg{\}}\, .
\end{equation}
%%%%%%%%%%%%%%%%%%%%%%%%%%%%%%%%%%%%%%%%%%%%%%%%%%%%%%%%%%%%%%%%%%%%%%%%%%
It was also found by \citet{RL21} that the critical redshift, $z_{c}$, appears to sensitively depend on the initial conditions,   
exhibiting a tendency to have a lower value for the case of the less amount of dark matter. 

\section{The effect of massive neutrinos on the splashback mass function}\label{sec:test}

We are going to numerically tackle the following two issues. First, is the analytic model reviewed in Section \ref{sec:review} still valid  for the $\nulcdm$ cosmology? 
Second, what effect does the presence of neutrinos has on the evolution of the diffusion coefficient, $\db (z)$?  
To address these issues, we utilize the snapshot data from the Cosmological Massive Neutrino Simulations ($\nus$) \citep{liu-etal18} 
for which the effect of neutrinos was incorporated into the background with the help of the analytic linear response approximation \citep{AB13}. 
In the periodic box of linear size $512\,h^{-1}$Mpc,  the $\nus$ contained a total of $N_{\rm dm}=1024^{3}$ DM particles with individual mass 
$10^{10}\,h^{-1}\,M_{\odot}$ \citep{liu-etal18}. 

The snapshot data from the $\nus$ are available only for two $\nulcdm$ models which share the same matter density parameter ($\Omega_{m}$) and same large-scale 
amplitude of the linear density power spectrum ($A_{s}$) but different total neutrino mass, $\num/\ev=0.0$ and $0.1$, respectively.  
Table \ref{tab:para} lists the values of $\Omega_{m}$, $A_{s}$ and $\sigma_{8}$, for the two models. Note that $\sigma_{8}$ has a slightly lower value in the presence of massive 
neutrinos due to the suppression of the small-scale powers caused by the neutrino free streaming \citep{LP14}. The catalogs of distinct halos and their subhalos identified by the 
Rockstar algorithm \citep{rockstar} in the snapshot data are also publicly available at the $\nus$ website \footnote{http://astronomy.nmsu.edu/aklypin/SUsimulations/MassiveNuS/}. 

We apply the publicly available SPARTA (Subhalo and PARticle Trajectory Analysis) algorithm \citep{sparta17} to a total of $45$ snapshots and rockstar catalogs in the redshift 
range of $0\le z\le 5.285$ from the $\nus$ for the two $\nulcdm$ cosmologies. The redshift interval, $\Delta z$, between the adjacent snapshots in the redshift range of $z\le 1$ 
is approximately $\delta z \approx 0.04$. 
The SPARTA basically inspects the orbits of all particles in each rockstar halo over redshifts to compute the distances to the apocenters 
of their first orbits, $r_{sp}$, and takes the average over them, $R^{mn}_{sp}$, to define the splashback radius of each halo. The SPARTA algorithm also provides several 
different definitions of the splashback radius, $R^{50}_{sp}$, $R^{70}_{sp}$, and $R^{90}_{sp}$, which correspond to the $50\%$, $70\%$, and $90\%$ percentiles of $r_{sp}$, 
respectively.  The analytic formula reviewed in Section \ref{sec:review} was proven to describe well the splashback mass function for the case of the $\Lambda$CDM 
cosmology, no matter which definition of $R_{sp}$ was used. In the current work, we exclusively use $R^{mn}_{sp}$ for the halo splashback radius, since the linear 
fit approximation to the evolution of the diffusion coefficient turned out to work best for the case of $R^{mn}_{sp}$ \citep{RL21}.

To determine the splashback mass function, we consider only the distinct halos that are not embedded within the splashback radii of any larger halos.  Note that the criterion for a 
distinct halo depends on the halo finder algorithm. For example, some rockstar object classified as a distinct halo by the spherical overdensity (SO) code could be classified as a 
subhalo by the MORIA code, a subrouine contained in the SPARTA algorithm, as its splashback radius can be overlapped with that of a neighbor larger halo. 
Selecting the distinct halos with splashback masses $M\ge 5\times 10^{12}\, \munit$, we split the range of $\ln M$ into short bins with equal 
length of $d\ln M$ and count the number of the selected halos, $dN$, whose logarithmic masses are in the range of $\left[\ln M, \ln M + d\ln M\right]$. 
Then, the splashback mass function is numerically determined as the ratio, $dN/d\ln M$, at each redshift, to which  the analytic model, 
Equations (\ref{eqn:massft})-(\ref{eqn:kappa}), is adjusted by varying $\db$.  Dividing the halo sample into eight subsamples of equal size, we also repeat the same 
calculation separately for each of the eight subsamples to obtain eight different splashback mass functions, $\{dN_{i}/d\ln M\}_{i=1}^{8}$. 
The one standard deviation scatter among $\{dN_{i}/d\ln M\}_{i=1}^{8}$ is then determined as the Jackknife errors in the original splashback mass function. 
As done in \citet{RL21}, the standard $\chi^{2}$-statistics is employed to find the best-fit value of $\db$, while the error in $\db$ is computed as the associated 
Fisher information. 

\begin{figure}
\begin{center}
\includegraphics[scale=0.7]{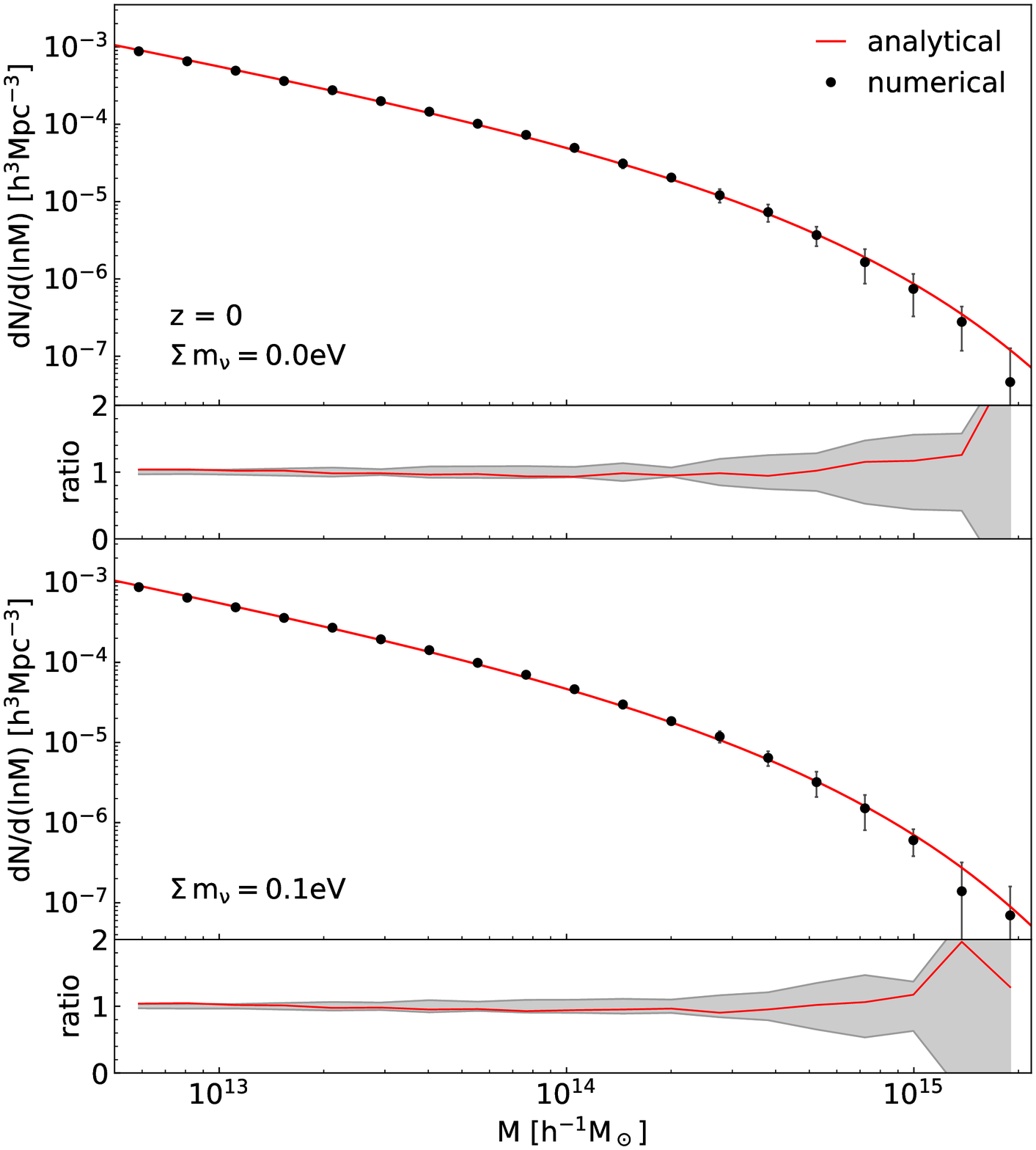}
\caption{Analytical formula of the splashback mass function (red solid line) with the diffusion coefficient 
$\db$ determined by adjusting the formula to the numerical results (black closed circles with errors) 
for the two models at $z=0$.}
\label{fig:massft0}
\end{center}
\end{figure}

Figure \ref{fig:massft0} plots the numerically obtained splashback mass function (black closed circles) with the Jackknife errors at $z=0$, and compares them with the 
analytic formula (red solid line) with the best-fit value of $\db$, for the two $\nulcdm$ models in the first and third from the top panels. 
The ratios of the numerical result to the analytical formula for each model (red solid line) with one standard deviation scatter (gray area) are also shown in the third 
and first from the bottom panels. 
As can be seen, for both of the $\nulcdm$ models, despite that it has only one parameter, the analytic formula excellently matches the numerical results in the wide mass 
range of $5\times 10^{12}\le M/(h^{-1}\,M_{\odot})\le 10^{15}$ at $z=0$. 

\begin{figure}
\begin{center}
\includegraphics[scale=0.7]{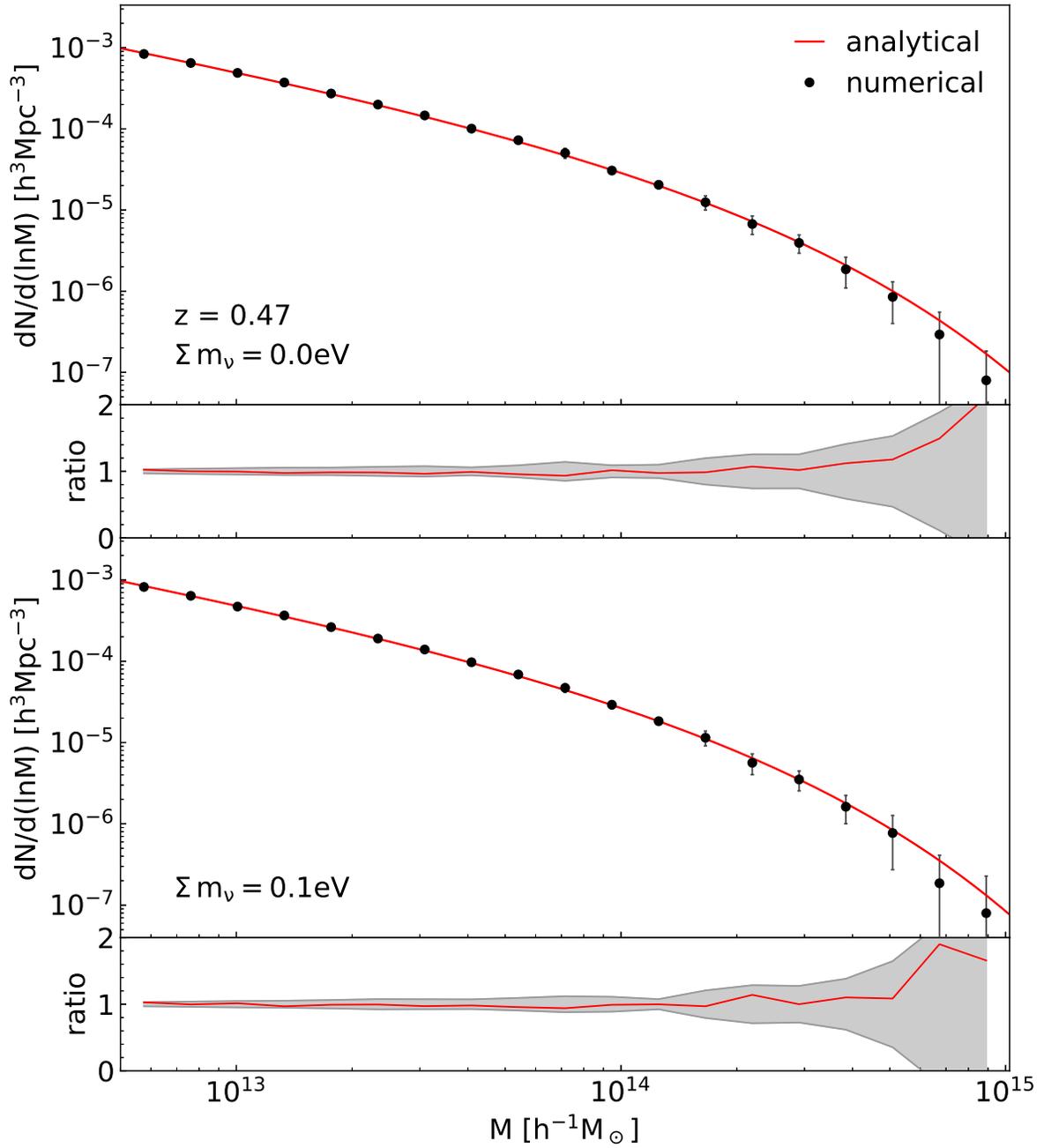}
\caption{Same as Figure \ref{fig:massft0} but at $z=0.47$.}
\label{fig:massft0.5}
\end{center}
\end{figure}

\begin{figure}
\begin{center}
\includegraphics[scale=0.7]{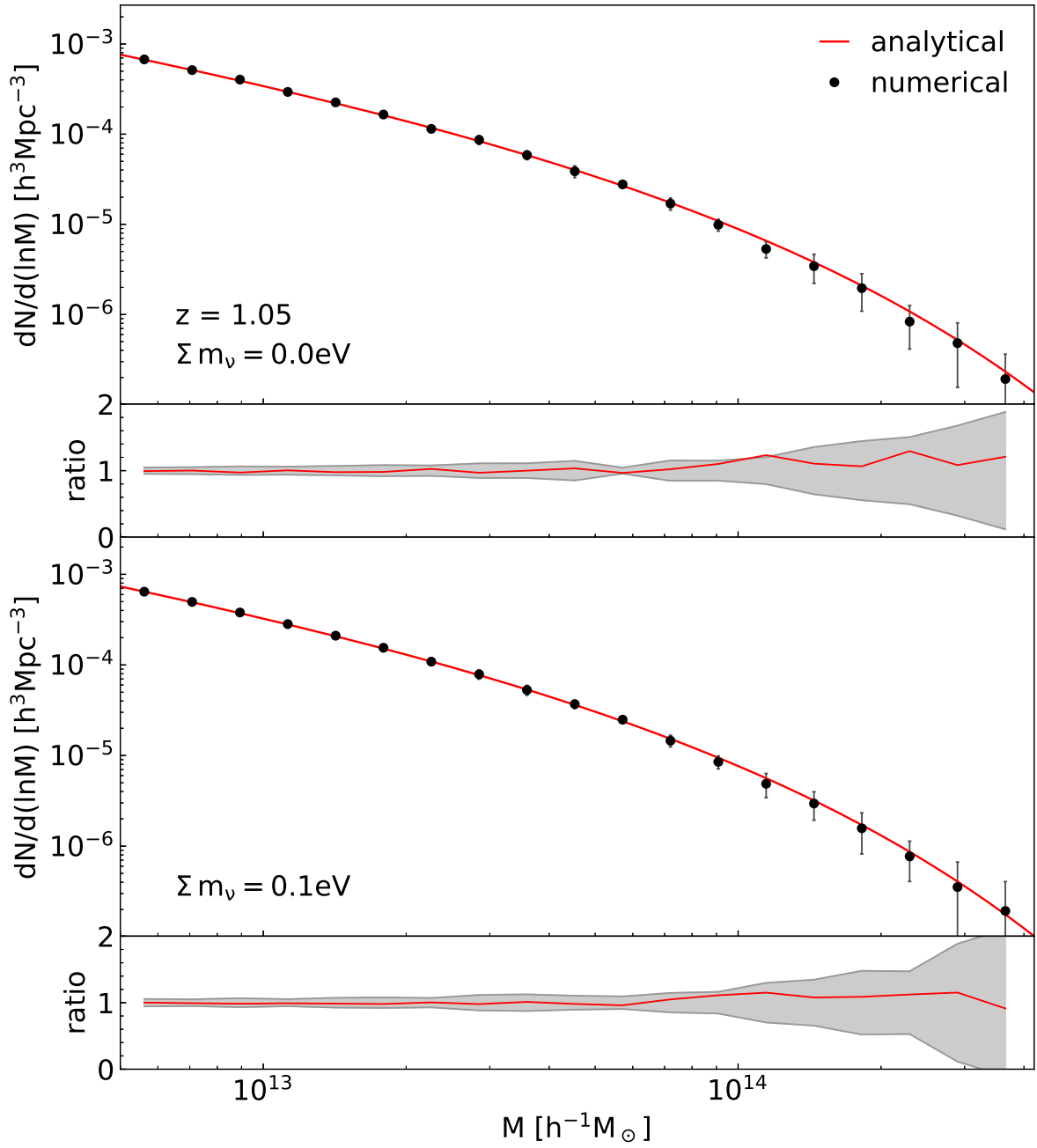}
\caption{Same as Figure \ref{fig:massft0} but at $z=1.05$.}
\label{fig:massft1}
\end{center}
\end{figure}

\begin{figure}
\begin{center}
\includegraphics[scale=0.7]{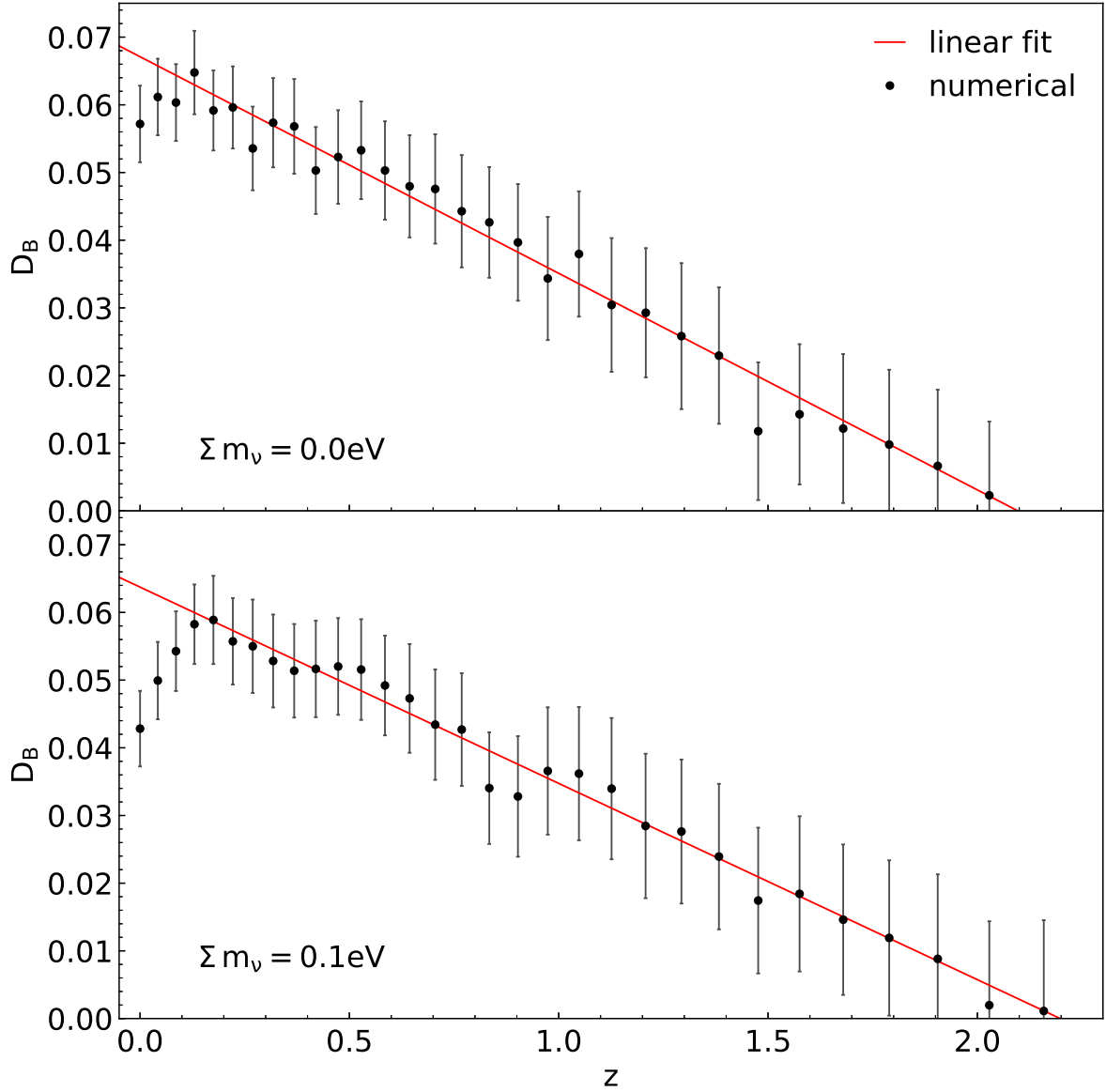}
\caption{Best-fit values of the diffusion coefficient of the splashback mass function versus redshifts. }
\label{fig:db}
\end{center}
\end{figure}

\begin{figure}
\begin{center}
\includegraphics[scale=0.7]{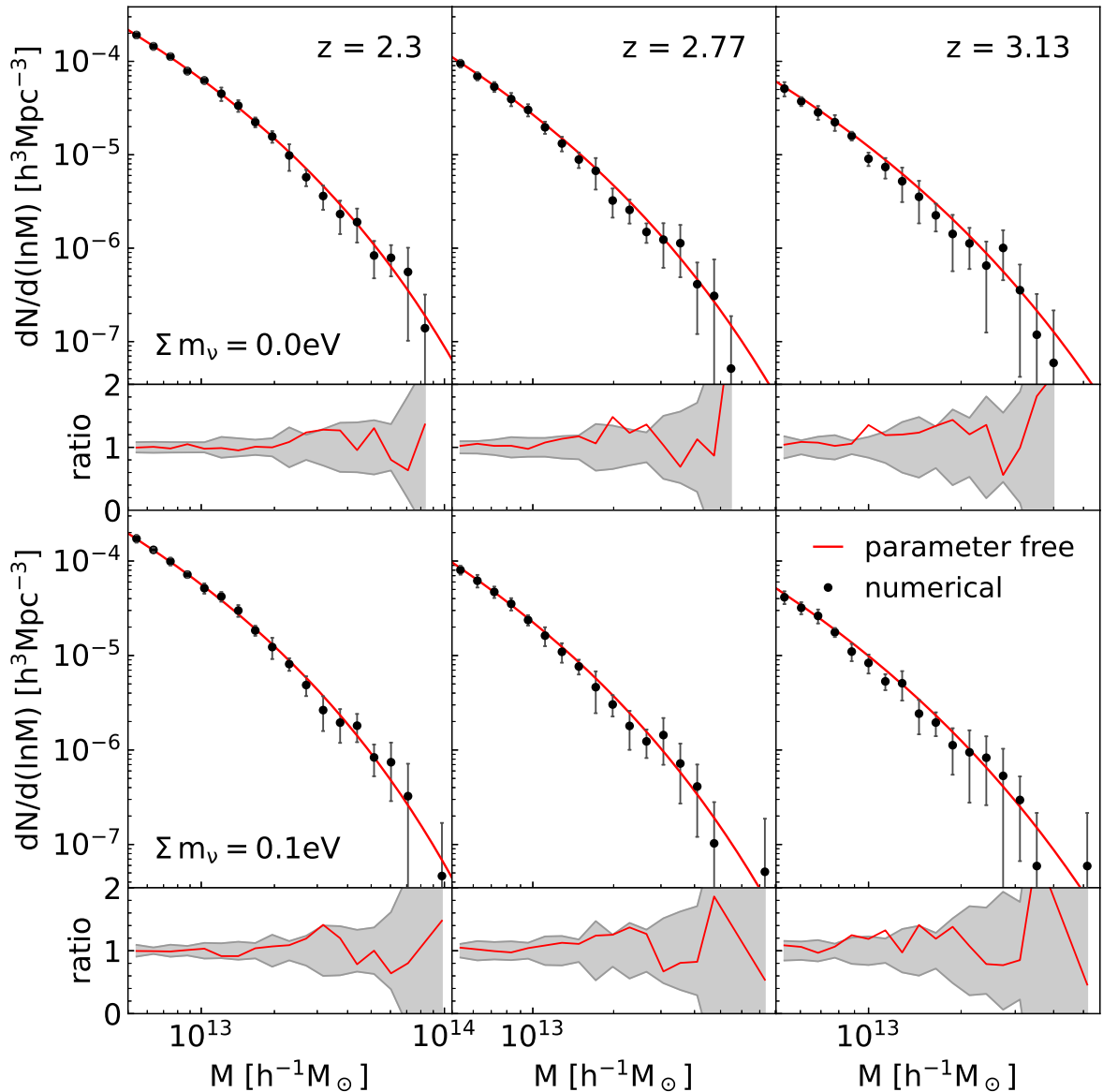}
\caption{Purely analytical parameter free formula of the the splashback mass function (red solid line) compared with 
the numerical results (black closed circles with errors) for the two models at three different redshifts higher 
than the critical redshift, $z>z_{c}$.}
\label{fig:free}
\end{center}
\end{figure}

Repeating the same calculation but at higher redshifts, we trace $\db (z)$. Figures \ref{fig:massft0.5}-\ref{fig:massft1} plot the same as Figure \ref{fig:massft0} but 
at $z=0.47$ and $1.05$, respectively,  demonstrating how well the analytic formula, Equations (\ref{eqn:massft})-(\ref{eqn:kappa}) with fixed $\delta_{sp}=1.52$, 
describes the splashback mass function even in the presence of neutrinos over a broad range of $z$.
Figure \ref{fig:db} shows how $\db (z)$ (black filled circles) changes with redshifts for the two cases of $\num$. As can be seen, for both of the cases,  $\db (z)$ decreases 
almost linearly with redshifts at $z\ge 0.2$, showing a trend of converging to zero at some critical redshift, $z_{c}$.  
If the diffusion coefficient vanishes, the analytic formula of the splashback mass function becomes parameter free. Figure \ref{fig:free} compares the numerical results 
at three edshifts beyond $z_{c}$ with the parameter-free model, Equation (\ref{eqn:para_free}).  Although the numerical results suffer from large Jackknife errors at these high 
redshifts due to the low number densities of massive cluster halos, the splashback mass functions agree quite well with the purely analytic model for both of the $\nulcdm$ models 
at $z>z_{c}$. These result are consistent with what \citet{RL21} found for the case of the $\Lambda$CDM cosmology, confirming the validity of 
Equations (\ref{eqn:massft})-(\ref{eqn:para_free}) even in the presence of neutrinos. 

As done in \citet{RL21}, approximating $\db (z)$ in the range of $0.2\le z \le z_{c}$ to a linear fit, $B(z_{c}-z)$, we determine the best values of $B$ and $z_{c}$ via the standard 
$\chi^{2}$-statistics, which are listed in Table \ref{tab:para} . As can be read, the two models differ in the best-fit values of $B$ and $z_{c}$,  the massive neutrino case yielding a 
substantially larger value of $z_{c}$ and significantly lower value of $B$ than the massless neutrino case. This result indicates that the presence of massive neutrinos has an effect 
of making $\db (z)$ more diffusive in this redshift range, slowing down the decrease of $\db (z)$ with $z$. In other words, in the presence of hot DM particles like massive 
neutrinos that have large free streaming lengths, the DM halos experience larger amount of disturbance from the surrounding,  which make it more ambiguous 
to precisely identify their physical boundaries. 

Unlike the result of \citet{RL21} that $\db (z)$ linearly decreases with redshifts in the whole range from $z=0$ to $z_{c}$, however, we find an anomalous behavior of $\db (z)$ 
at $z<0.2$ in the presence of neutrinos: it increases with $z$, deviating from the linear fit at $z<0.2$, as can be seen in Figure \ref{fig:db}. 
We also note that the massive neutrino case exhibits a much more conspicuous deviation of $\db (z)$ from the linear fit than the massless case at these low redshifts. 
We suspect that it should be caused or at least linked closely with the inherent limitation of the SPARTA code in locating the splashback boundaries of DM halos at $z\le 0.13$ 
where the unknown future orbits of infalling particles make it impossible to determine particle apocenters (private communication with B. Diemer 2022). 

\section{Summary and Conclusion}\label{sec:sum}

We have numerically investigated the effect of neutrinos on the splashback mass functions of DM halos in the mass range of $0.5\le M/(10^{13}\,h^{-1}\,M_{\odot})\le 10^{2}$ 
by applying the SPARTA code \citep{sparta17} to the snapshots of $\nus$ \citep{liu-etal18} for two different $\nulcdm$ cosmologies with total neutrino mass of $\num=0.0\ev$ 
and $0.1\ev$. Comparing the numerically obtained splashback mass functions to the analytic single parameter formula proposed by \citet{RL21} at various redshifts, we have 
trailed the evolution of the single parameter, dubbed the diffusion coefficient $\db (z)$, in the range of $0\le z\le 3$. 
It has turned out that the analytic formula with fixed splashback density threshold $\delta_{sp}=1.52$ validly describes the splashback mass function for both of the 
$\nulcdm$ models at redshifts up to $z\approx 3$ and that the diffusion coefficient decreases almost linearly with redshifts at $z\ge 0.2$, evanescing at some critical 
redshift, which are all in consistent with the claim of \citet{RL21}. 

Fitting $\db (z)$ at $z\ge 0.2$ to a linear scaling relation of $B(z-z_{c})$ for both of the $\nulcdm$ models, we have determined the best-fit values of the slope, $B$, 
and critical redshift, $z_{c}$, and find that they sensitively depend on the total neutrino mass. The $\nulcdm$ model with $\num=0.1\ev$ yields a significantly lower value 
of $B$ and a substantially higher value of $z_{c}$ than the other model with $\num=0.1\ev$, which indicates that the diffusion coefficient decreases more slowly with $z$ 
in the presence of massive neutrinos.  Since the diffusion coefficient of the analytic formula is a measure of the stochasticity of the splashback density threshold, this result 
implies that the presence of hot DM particles like massive neutrinos has an effect of disturbing more severely the splashback boundaries and making the halo identification 
more ambiguous. The dependence of $B$ and $z_{c}$ on $\num$ also implies that the diffusion coefficient of the splashback mass function can in principle be used as  
a probe of the total neutrino mass. 

A failure of the linear fit approximation to $\db (z)$ has been witnessed for both of the models at $z<0.2$ when the diffusion coefficient exhibit increment rather than 
decrement with $z$, an anomalous tendency never detected in the absence of neutrinos for which case the linear fit always matches the diffusion coefficient in 
the entire range of $0\le z< 3$.  Given the fundamental limitation of the SPARTA code at $z\le 0.13$ that it is incapable of 
locating the apocenters of infalling DM particles whose future orbits are known (private communication with B. Diemer 2022), we suspect that the anomalous behavior 
of $\db (z)$ witnessed at $z<0.2$ should not be a real one caused by the presence of massive neutrinos but a spurious one caused by the inherent flaw of the SPARTA 
code at low redshifts. 

It is, however, worth discussing a caveat that our conclusion is subject to. The $\nus$ employed the analytic linear response approximation \citep{AB13} to mimic the presence 
of  massive neutrinos rather than directly including the massive neutrinos as component particles. Although this approximation has been known to work quite well 
provided that $\num\le 0.3\ev$ \citep{bir-etal18}, it is not guaranteed to accommodate all possible effects that the massive neutrinos would have on the nonlinear evolution 
of cosmic web. It will require a more rigorous and accurate treatment of the neutrino effects to investigate the true behavior $\db (z)$ in the nonlinear regime and its dependence 
on the total neutrino mass.  It will be also quite desirable to have an analytic model for $\db (z)$ derived from the first principles, with which one can quantitatively probe the nature 
of dark matter including massive neutrinos. Our future work is in this direction. 

\acknowledgments

We are very grateful to B. Diemer for his many helps with the application of the SPARTA code to the \nus data from the Columbia Lensing group as well as 
his enlightening comments, which helped us avoid a misinterpretation of our results and significantly improve the original manuscript.
We thank the Columbia Lensing group for making their suite of simulated maps available at the website (http://columbialensing.org), and NSF for supporting 
the creation of those maps through grant AST-1210877 and XSEDE allocation AST-140041. We thank the New Mexico State University (USA) and Instituto de 
Astrofisica de Andalucia CSIC (Spain) for hosting the Skies \& Universes site for cosmological simulation products. We thank J.Liu for providing us with the 
$\nus$ snapshot data. We acknowledge the support by Basic Science Research Program through the National Research Foundation (NRF) of 
Korea funded by the Ministry of Education (No.2019R1A2C1083855).

\clearpage
%%%%%%%%%%%%%%%%%%%%%%%%%%%%%%%%%%%%%%%%%%%%%%%%%%
%%%%%%%%%%%%%%%%%%%%%%%%%%%%%%%%%%%%%%%%%%%%%%%%%%
\clearpage
%%%%%%%%%%%%%%%%%%%%%%%%%%%%%%%%%%%%%%%%%%%%%%%%%%
%%%%%%%%%%%%%%%%%%%%%%%%%%%%%%%%%%%%%%%%%%%%%%%%%%
\clearpage
%%%%%%%%%%%%%%%%%%%%%%%%%%%%%%%%%%%%%%%%%%%%%%%%%%
%%%%%%%%%%%%%%%%%%%%%%%%%%%%%%%%%%%%%%%%%%%%%%%%%%\
\clearpage
%%%%%%%%%%%%%%%%%%%%%%%%%%%%%%%%%%%%%%%%%%%%%%%%%%
%%%%%%%%%%%%%%%%%%%%%%%%%%%%%%%%%%%%%%%%%%%%%%%%%%
\clearpage
%%%%%%%%%%%%%%%%%%%%%%%%%%%%%%%%%%%%%%%%%%%%%%%%%%
%%%%%%%%%%%%%%%%%%%%%%%%%%%%%%%%%%%%%%%%%%%%%%%%%%
%%%%%%%%%%%%%%%%%%%%%%%%%%%%%%%%%%%%%%%%%%%%%%%%%%
\clearpage
\begin{deluxetable}{cccccc}
\tablewidth{0pt}
\setlength{\tabcolsep}{5mm}
\tablecaption{Best-fit values of the critical redshift for the massless and massive neutrino cases.}
\tablehead{ $\num$ & $\Omega_m$ & $A_{s}$ & $\sigma_8$ &  $B$ & $z_c$ \\
$[\ev]$ & & $(10^{-9})$ & &  & } 
\startdata
$0.0$ & $0.3$	& $2.1$ & $0.85$ & $-0.032\pm 0.001$ & $2.097\pm 0.047$ \\
$0.1$ & $0.3$	& $2.1$ & $0.83$ & $-0.029\pm 0.001$ & $2.198\pm 0.044$ \\
\enddata
\label{tab:para}
\end{deluxetable}
%%%%%%%%%%%%%%%%%%%%%%%%%%%%%%%%%%%%%%%%%%%%%%%%%%

\end{document}